\documentclass[pdflatex]{article}
\usepackage{spconf,amsmath}
\usepackage{graphicx}
\usepackage{bm}
\usepackage{url}
\usepackage{amssymb}
\usepackage{cite}

\newcommand{\argmin}{\mathop{\rm arg~min}\limits}

\makeatletter
\def\Hline{
  \noalign{\ifnum0=`}\fi\hrule \@height 2.\arrayrulewidth \futurelet
  \reserved@a\@xhline}
\makeatother

\title{Trainable Adaptive Window Switching for Speech Enhancement}
\name{Yuma Koizumi$^{\dagger}$, Noboru Harada$^{\dagger}$, and Yoichi Haneda$^{\ddagger}$}
\address{
\normalsize{$^{\dagger}$: NTT Media Intelligence Laboratories, Tokyo, Japan}\\
\normalsize{$^{\ddagger}$: The University of Electro-Communications, Tokyo, Japan}
}

\begin{document}
\ninept
\maketitle

\begin{abstract}
This study proposes a trainable adaptive window switching (AWS) method and apply it to a deep-neural-network (DNN) for speech enhancement in the modified discrete cosine transform domain.
Time-frequency (T-F) mask processing in the short-time Fourier transform (STFT)-domain is a typical speech enhancement method. 
To recover the target signal precisely, DNN-based short-time frequency transforms have recently been investigated and used instead of the STFT. 
However, since such a fixed-resolution short-time frequency transform method has a T-F resolution problem based on the uncertainty principle, not only the short-time frequency transform but also the length of the windowing function should be optimized. 
To overcome this problem, we incorporate AWS into the speech enhancement procedure, and the windowing function of each time-frame is manipulated using a DNN depending on the input signal. 
We confirmed that the proposed method achieved a higher signal-to-distortion ratio than conventional speech enhancement methods in fixed-resolution frequency domains.
\end{abstract}

\begin{keywords}
Speech enhancement, trainable time-frequency representation, adaptive window switching, MDCT.
\end{keywords}

\vspace{-0pt}


\section{Introduction}
\label{sec:intro}

Speech enhancement is used to recover the target speech from a noisy observed signal. 
A recent advancement in this area is the use of deep learning to estimate a time-frequency (T-F) mask 
\cite{Wang_2018,Erdogan_2015,Hershey_2016,Kolbak_2017,Koizumi_ICASSP_2017,Koizumi_TASL_2018}; 
a T-F mask is estimated using a deep-neural-network (DNN) and applied to T-F represented observation, then the estimated signal is re-synthesized using the inverse transform. 
Traditionally, the short-time Fourier transform (STFT) and a real-valued T-F mask is used as a T-F transform and its T-F mask, respectively. 
This means most algorithms only manipulate the magnitude; thus, the performance upper bound is limited by the noisy phase. 
To overcome this limit, phase-reconstruction methods, 
including 
complex-valued T-F mask estimation \cite{Will_cIRM_2016}, 
consistency-based methods \cite{GL_1984,Yatabe_2018}, 
model-based methods \cite{Wakabayashi_2018,Masuyama_2018}, 
and DNN-based phase estimation \cite{PhaseNet,GAN_phase,Phasebook,DeGLI}, have been investigated.

In contrast to the phase-reconstruction methods, the use of another T-F transforms have also been investigated. 
Using a real-valued T-F transform, such as the modified discrete cosine transform (MDCT) \cite{MDCT_01}, 
enables us to avoid dealing with phase prediction \cite{Kuech_2007}, and 
we have reported that a DNN for estimating a T-F mask in the MDCT domain can be trained by extending DNN-based source enhancement to end-to-end manner \cite{Koizumi_ICASSP_2018}. 
More recently, trainable T-F transforms have been investigated such as 
auto-encoder transform \cite{Venkataramani_2017,Venkataramani_2018}, 
STFT convolution \cite{Wichern_2018}, TasNet \cite{Luo_2018}, and 
the use of the warped filter bank frame \cite{Takeuchi_2019}. 
Here, ``trainable'' means that the parameters of transform can be trained for minimizing an objective function. 
These studies suggest the existence of a more suitable basis-domain than the STFT-domain for speech enhancement. 

\begin{figure}[ttt]
  \centering
\includegraphics[width=85mm,clip]{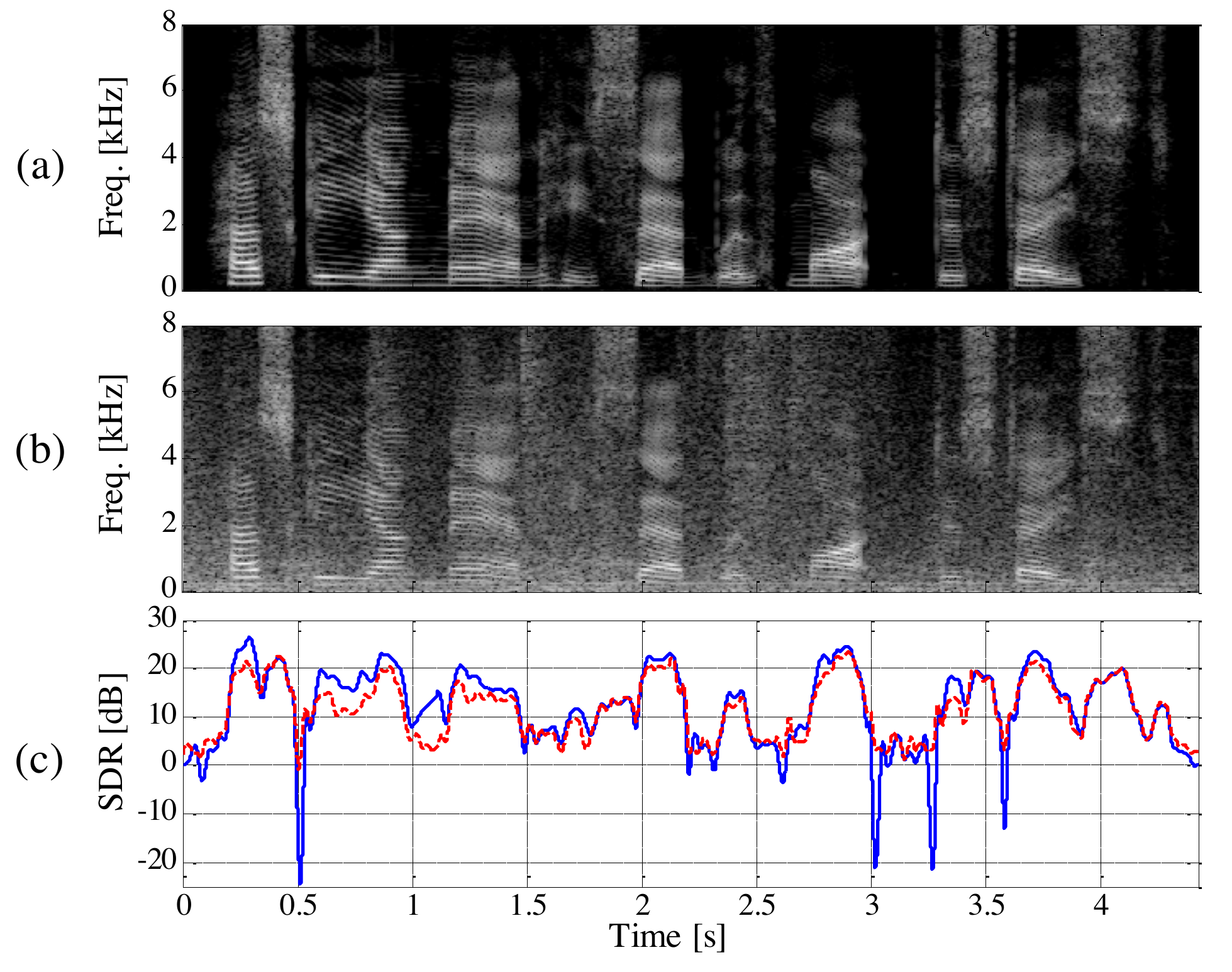}  
  \caption{
Spectrograms of (a) clean and (b) noisy speech, and (c) segmental SDRs with $[0, 1]$ truncated oracle T-F mask in MDCT-domain.
Blue line and red dotted line denote segmental SDRs when $L=1024$ and $L=128$, respectively.
}
  \label{fig:flow}
\end{figure}

Another problem of T-F analysis in audio signal processing is the T-F resolution tradeoff; a fixed-resolution short-time frequency transform has a T-F resolution problem based on the uncertainty principle.
Figure \ref{fig:flow} shows an example of this problem in speech enhancement. 
The length of the windowing function $L$ relates to the resolution of both time and frequency components; 
a long $L$ results in better frequency resolution but poor time resolution, and vice versa. 
Thus, although a long $L$ results in a higher segmental signal-to-distortion ratio (SDR) in the stationary phoneme intervals, 
it also results in a worse segmental SDR at the change points of phonemes and/or consonant intervals. 
Thus, to recover the target signal more precisely, 
not only the basis functions of the short-time frequency transform but also $L$ should be manipulated depending on the characteristics of each time-frame.

We propose a trainable adaptive window switching (AWS) method and apply it to the MDCT-domain speech enhancement.
In AWS \cite{Mochizuki_1994,AWS}, $L$ is manipulated depending on the characteristics of each segment, and $L$ is operated by a binary variable that denotes whether the target frame should be analyzed using a long or short window. 
Thus, since the unknown parameter of AWS is the binary variable, the proposed method estimates this variable by using a DNN, and both a binary-decision DNN and mask-estimation DNN are simultaneously trained to minimize the same objective function.

\section{Conventional method}
\label{sec:conv}

\subsection{General form of T-F mask processing}

Let us consider that the $K$ samples of time-domain observation $\bm{x} = (x_1, x_2,...,x_K)^{\top}$ is a mixture of a target $\bm{s}$ and noise $\bm{n}$ as 
\begin{equation}
\bm{x} = \bm{s} + \bm{n},
\end{equation}
where $\top$ denotes the transposition. 
The goal with speech enhancement can be formulated as recovering an estimate of $\bm{s}$ as $\hat{\bm{s}}$ from $\bm{x}$. 
In T-F mask processing, $\hat{\bm{s}}$ can be estimated using two functions; 
a T-F transform function $\mathcal{P} : \bm{x} \mapsto \bm{X}$ and T-F mask estimator $\mathcal{M}_{\theta}$ with parameter $\theta$. 
Here, $\bm{X}$ is a T-F representation of $\bm{x}$, and $\mathcal{M}_{\theta}$ outputs a T-F mask with the same size as $\bm{X}$. 
Thus, T-F mask processing can be generally written as
\begin{align}
\hat{\bm{s}} = \mathcal{P}^{-1} \left[  \mathcal{M}_{\theta} \left( \bm{\phi} \right) \odot \mathcal{P}[ \bm{x} ]  \right], 
\label{eq:general_TFmask}
\end{align}
where $\mathcal{P}^{-1}$ is the (pseudo-)inverse transform of $\mathcal{P}$, $\bm{\phi}$ is an acoustic feature extracted from $\bm{x}$, and $\odot$ denotes the element-wise product. 
In most cases, $\mathcal{P}$ is taken to be the STFT, and $\mathcal{M}_{\theta}$ returns a real-valued T-F mask. These values are constrained to lie between 0 to 1. 
Recently, $\mathcal{M}_{\theta}$ has been implemented using a DNN, 
and $\theta$ has been trained to minimize an objective function $\mathcal{J}_{\theta}$ by using the gradient method.

A problem with T-F mask processing in the STFT-domain is that a real-valued T-F mask only manipulates the magnitude; thus, the upper bound of speech enhancement performance is limited by the noisy phase. 
There are roughly two solutions, {\it i.e.}, the use of a phase-reconstruction methods \cite{Will_cIRM_2016,GL_1984,Yatabe_2018,Wakabayashi_2018,Masuyama_2018,PhaseNet,GAN_phase,Phasebook} 
or another T-F transform \cite{Koizumi_ICASSP_2018,Venkataramani_2017,Venkataramani_2018,Wichern_2018,Luo_2018}. 
In this study, we focus on the later, and in the next section, we briefly describe speech enhancement in the MDCT-domain \cite{Koizumi_ICASSP_2018}.

\subsection{T-F mask processing in the MDCT-domain}

First, we separate $\bm{x}$ into $T$ short-time signals of length $\mathcal{L} = L/2$ without overlap, where an even number $L$ is the length of the windowing function. 
Then, the $t$-th separated signal is written as 
\begin{align}
\bm{\mathrm{x}}_t := ( x_{\mathcal{L}(t-1)+ 1}, x_{\mathcal{L}(t-1)+ 2}, ... , x_{\mathcal{L}(t-1)+ \mathcal{L}} )^{\top}.
\label{eq:frame_time_x}
\end{align}
Then, the MDCT and its inverse can be written as
\begin{align}
\bm{\mathrm{X}}_t^{C} = \bm{ \mathrm{M} } \left[
\begin{matrix}
\bm{\mathrm{x}}_{t-1}\\
\bm{\mathrm{x}}_{t}
\end{matrix}
\right], \;\;
\left[
\begin{matrix}
\bm{\mathrm{x}}_{t}^{(C1)}\\
\bm{\mathrm{x}}_{t}^{(C2)}
\end{matrix}
\right]
=
\bm{ \mathrm{M} }^{\top} \bm{\mathrm{X}}_t^{C},
\label{eq:IMDCT}
\end{align}
respectively. 
Here, $\bm{\mathrm{X}}_t^{C} := \left( X_{t, 1}^{C}, ..., X_{t, \mathcal{L}}^{C} \right)^{\top}$ are 
MDCT coefficients and $\bm{ \mathrm{M} } = \bm{\mathrm{C}} \bm{\mathrm{W}} \in \mathbb{R}^{\mathcal{L} \times L}$ is the analysis matrix. 
The matrices $\bm{\mathrm{C}} \in \mathbb{R}^{\mathcal{L} \times L}$ and $\bm{\mathrm{W}} \in \mathbb{R}^{L \times L}$ are the MDCT matrix and a diagonal matrix for windowing, respectively. 
In the MDCT, the analysis/synthesis windowing function must satisfy the Princen-Bradley condition \cite{MDCT_01}, and the sine-window is typically used. 
Since $\bm{\mathrm{C}}$ is not a square matrix, it does not have the inverse. 
Thus, $\bm{\mathrm{x}}_{t} \neq \bm{\mathrm{x}}_{t}^{(C2)}$ and $\bm{\mathrm{x}}_{t}^{(C2)}$ include time-domain aliasing. 
In the MDCT, this aliasing can be canceled by overlap-add as
\begin{align}
\bm{\mathrm{x}}_{t} 
= \bm{\mathrm{x}}_{t}^{(C2)} + \bm{\mathrm{x}}_{t+1}^{(C1)}.
\label{eq:tdac}
\end{align}
Since $\mathcal{P}$ and $\mathcal{P}^{-1}$ are defined with (\ref{eq:IMDCT}) and (\ref{eq:tdac}), 
generalized T-F mask processing (\ref{eq:general_TFmask}) is possible in the MDCT-domain as follows:
\begin{align}
\left[
\begin{matrix}
\hat{\bm{\mathrm{s}}}_{t}^{(C1)}\\
\hat{\bm{\mathrm{s}}}_{t}^{(C2)}
\end{matrix}
\right] =
\bm{ \mathrm{M} }^{\top} 
\left( \mathcal{M}_{\theta} \left( \bm{\phi}_t \right) \odot
\bm{ \mathrm{M} } \left[
\begin{matrix}
\bm{\mathrm{x}}_{t-1}\\
\bm{\mathrm{x}}_{t}
\end{matrix}
\right]
\right),
\label{eq:MDCT_DNNenh}
\end{align}
and $\hat{\bm{s}}$ is calculated by adding these outputs as
\begin{equation}
\hat{\bm{\mathrm{s}}}_{t} = \hat{\bm{\mathrm{s}}}_{t}^{(C2)} + \hat{\bm{\mathrm{s}}}_{t+1}^{(C1)}.
\label{eq:mdctout}
\end{equation}

\subsection{Adaptive window switching in the MDCT-domain}
\label{sec:vanilla}

\begin{figure}[ttt]
  \centering
\includegraphics[width=85mm,clip]{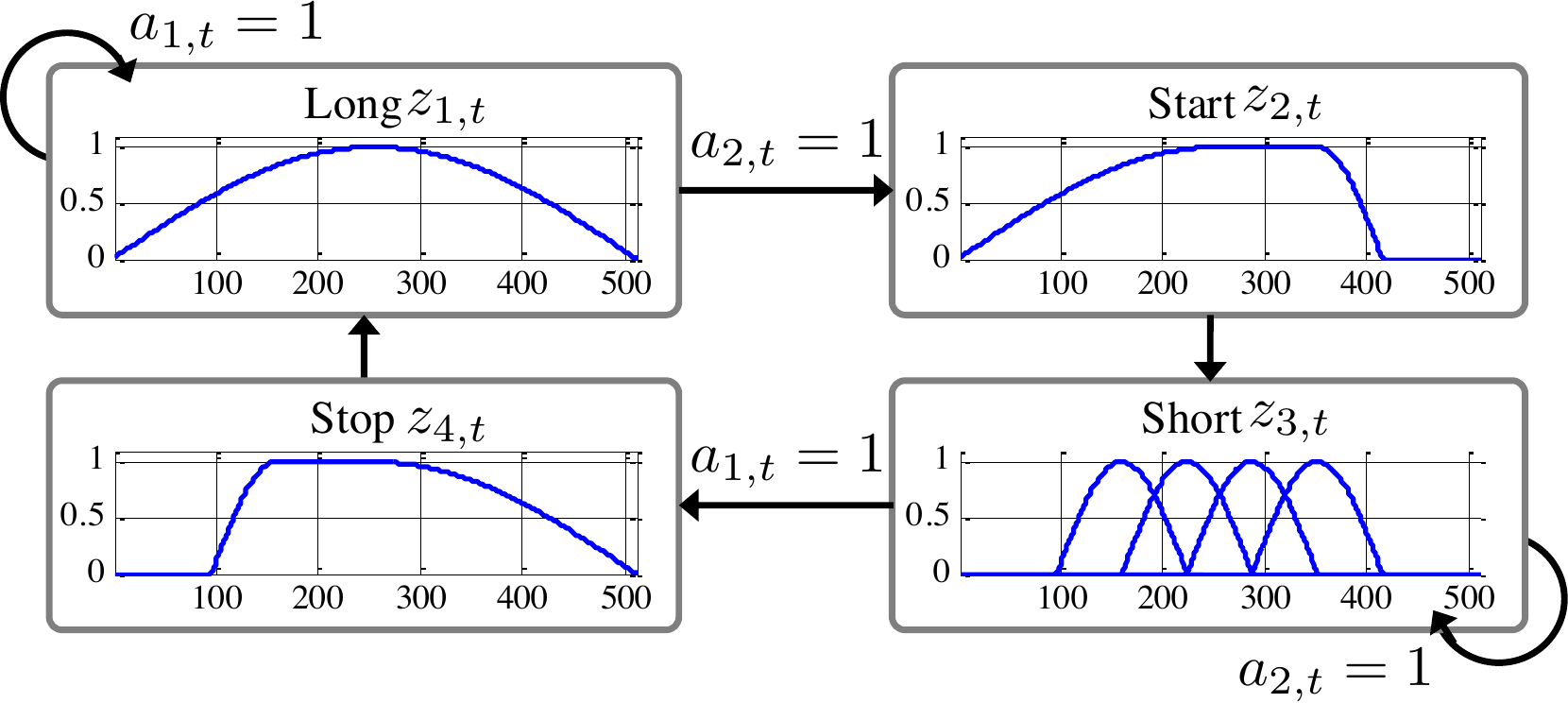}  
  \caption{
Example of window-switching rule when $L_{\mbox{\tiny long}}=512$ and $L_{\mbox{\tiny short}}=128$. 
X-axis of each figure denotes sample index. 
To guarantee PR property, transition windows ``start'' and ``stop'' are used, and window switching is manipulated using one-hot-vector $\bm{a}_t = (a_{1,t}, a_{2,t})$.
}
  \label{fig:faws}
\end{figure}

Although T-F mask processing is powerful for speech enhancement, it may have a T-F resolution problem, as shown in Fig. \ref{fig:flow}. 
The AWS in the MDCT-domain \cite{Mochizuki_1994,AWS} overcomes the T-F resolution problem without losing the perfect-reconstruction (PR) property by switching the four types of windows labeled ``long'', ``start'', ``short'', and ``stop''. 
A ``long'' window with length $L_{\mbox{\tiny long}}$ is used when the signal spectrum remains stationary or varies slowly over time. 
When the signal changes rapidly, a ``short'' window with length $L_{\mbox{\tiny short}}$ is used. 
The transition windows ``start'' and ``stop'' are used to change windows without losing the PR property; the start window is used in a transition from long to short and vice versa. 
This transition is manipulated using a one-hot-vector $\bm{a}_t = (a_{1,t}, a_{2,t})$. If $a_{1,t} = 1$ or $a_{2,t} = 1$, the window is changed to be ``long'' or ``short'', respectively, as shown in Fig \ref{fig:faws}. 
In the audio-coding area, $\bm{a}_t$ is determined based on a psycho-acoustics model \cite{mp3}. 

\section{Proposed method}
\label{sec:prop}

\subsection{Trainable adaptive window switching}
\label{sec:prop_aws}

\begin{figure}[ttt]
  \centering
\includegraphics[width=85mm,clip]{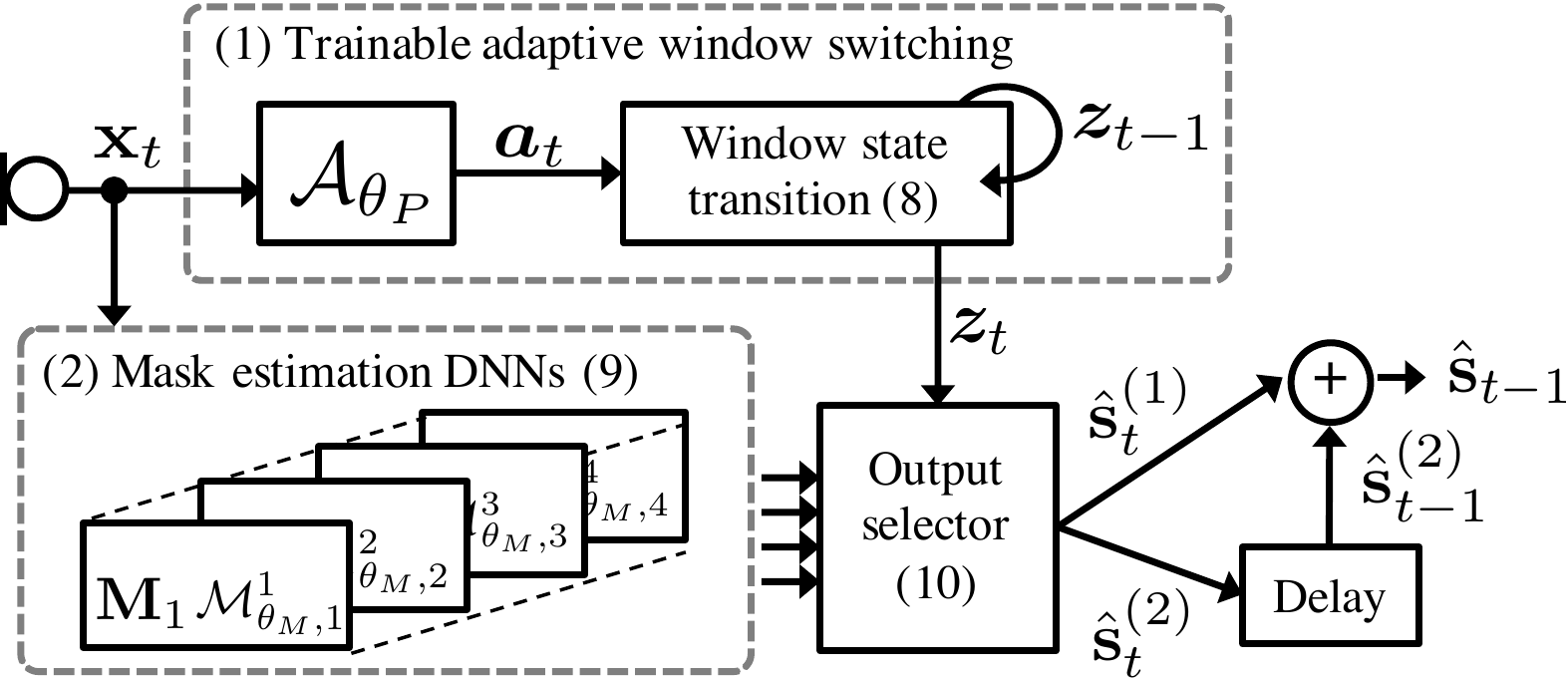}  
  \caption{Speech enhancement flowchart of proposed method.}
  \label{fig:proposed}
\end{figure}

Since fixed-resolution T-F transform connotes the T-F resolution trade-off, as shown in Fig \ref{fig:flow}, 
for speech enhancement, not only the short-time frequency transform but also the window lengths should be trained to change $L$. 
Thus, we propose a speech enhancement method with a trainable AWS, as shown in Fig. \ref{fig:proposed}.

First, we generalize trainable T-F transform. ``Trainable'' means that a T-F analysis function $\mathcal{P}$ is parameterized by $\theta_P$, and we can train $\theta_P$ to minimize an objective function. 
Thus, in contrast to (\ref{eq:general_TFmask}), generalized T-F mask processing with a trainable T-F transform can be written as $\hat{\bm{s}} = \mathcal{P}^{-1}_{\theta_P} \left[ \mathcal{M}_{\theta_M} \left( \bm{\phi} \right) \odot \mathcal{P}_{\theta_P}[ \bm{x} ] \right].$ 
In AWS, the four types of windows are switched using $\bm{a}_t = (a_{1,t}, a_{2,t})$; thus, we estimate $\bm{a}_t$ by using a DNN $\mathcal{A}_{\theta_P}$ and incorporated into a T-F analysis function. 
Since $\bm{a}_t$ is a one-hot-vector, the sigmoid or softmax activation is not suitable for estimating $\bm{a}_t$. 
To use the back-propagation algorithm, logical operators, such as ``{\it switch}'' and/or ``{\it if}'', are not also suitable because the output signal needs to be differentiable w.r.t. $\theta_P$. 
Thus, as an implementation, we use the Gumbel-softmax activation \cite{gumbel-softmax} to obtain $\bm{a}_t = \mathcal{G} \left( \mathcal{A}_{\theta_P} [ \bm{\mathrm{x}}_t ], \tau \right),$ where $\mathcal{G}$ is the Gumbel-softmax activation and $\tau$ is the sofmax temprature.

Then, a one-hot-vector $\bm{z}_t = (z_{1,t}, z_{2,t}, z_{3,t}, z_{4,t})^{\top}$, which denotes the selected window at time-frame $t$, can be calculated by the following recursive formula as
\begin{align}
z_{k,t}	=z_{k, t-1} + \sum_{i=1}^{2} \sum_{j=1}^{4} a_{i,t} z_{j, t-1}  \bm{Q}_{ k, j, i},
\label{eq:aws_equation}
\end{align}
where $z_{1, t}=1$, $z_{2, t}=1$, $z_{3, t}=1$, and $z_{4, t}=1$ denote the selected window at $t$ as ``long'', ``start'', ``short'' and ``stop'', respectively. 
The matrices $\bm{Q}_{:,:,i}$ are the following state-transition matrices:
\begin{align}
\nonumber
\bm{Q}_{:,:,1} = 
 \begin{bmatrix}
 0		& 0		& 0 		&	1\\
 0		&-1		& 0 		&	0\\
 0		& 1		&-1 		&	0\\
 0		& 0		& 1 		&	-1\\
 \end{bmatrix}, \;
\bm{Q}_{:,:,2} = 
 \begin{bmatrix}
 -1		& 0		& 0 		&	1\\
 1		&-1		& 0 		&	0\\
 0		& 1		& 0	 	&	0\\
 0		& 0		& 0 		&	-1\\
 \end{bmatrix}.
\end{align}
As an example of (\ref{eq:aws_equation}), 
when $a_{1,t} = 1$ and the window of $t-1$ is ``short'' $\bm{z}_{t-1} = (0, 0, 1, 0)^{\top}$, the $j=3$rd column of $\bm{Q}_{:,:,1}$ is added to $\bm{z}_{t-1}$.
Namely, $\bm{z}_{t} = \bm{z}_{t-1} + \bm{Q}_{:,3,1} = (0, 0, 0, 1)^{\top}$. Thus, the window at $t$ is ``stop''.

Since the windowing function at $t$ is selected, the output signal can be obtained with four MDCT analysis matrices $\bm{ \mathrm{M} }_j$ and DNN-based T-F mask estimators $\mathcal{M}^j_{\theta_{M,j}}$ corresponding to the $j$-th window. 
The implementation of $\bm{ \mathrm{M}}_j$ is described in the next section. 
First, the output signal of the $j$-th window is calculated as
\begin{align}
 \begin{bmatrix}
 \hat{ \bm{ \mathrm{s} } }_{j, t}^{(C1)}\\
 \hat{ \bm{ \mathrm{s} } }_{j, t}^{(C2)}
 \end{bmatrix} = 
 \bm{ \mathrm{M} }_{j}^{\top}
 \left( 
 \mathcal{M}^j_{\theta_{M,j}}( \bm{\phi}_{j,t} ) 
 \odot
 \bm{ \mathrm{M} }_{j} 
 \begin{bmatrix}
 \bm{ \mathrm{x} }_{t-1}^{l}\\
 \bm{ \mathrm{x} }_{t}^{l}
 \end{bmatrix} 
 \right) , 
\end{align}
where $\bm{\phi}_{j,t}$ is the input vector for the $j$-th window at $t$. 
Then, since $\bm{z}_t$ is a one-hot-vector, the output signal can be obtained as a 
$z_{j,t}$-weighted sum of $\hat{ \bm{ \mathrm{s} } }_{j, t}^{(C1)}$ and $\hat{ \bm{ \mathrm{s} } }_{j, t}^{(C2)}$ as follows:
\begin{align}
 \hat{ \bm{ \mathrm{s} } }_{t} &= 
\sum_{j=1}^{4}  z_{j, t}\hat{ \bm{ \mathrm{s} } }_{j, t}^{(C2)}
+
\sum_{j=1}^{4} z_{j, t+1}\hat{ \bm{ \mathrm{s} } }_{j, t+1}^{(C1)} 
\label{eq:prop_out_eq}.
\end{align}

\subsection{Implementation of analysis matrices}
\label{sec:implementation}

\begin{figure}[ttt]
  \centering
\includegraphics[width=80mm,clip]{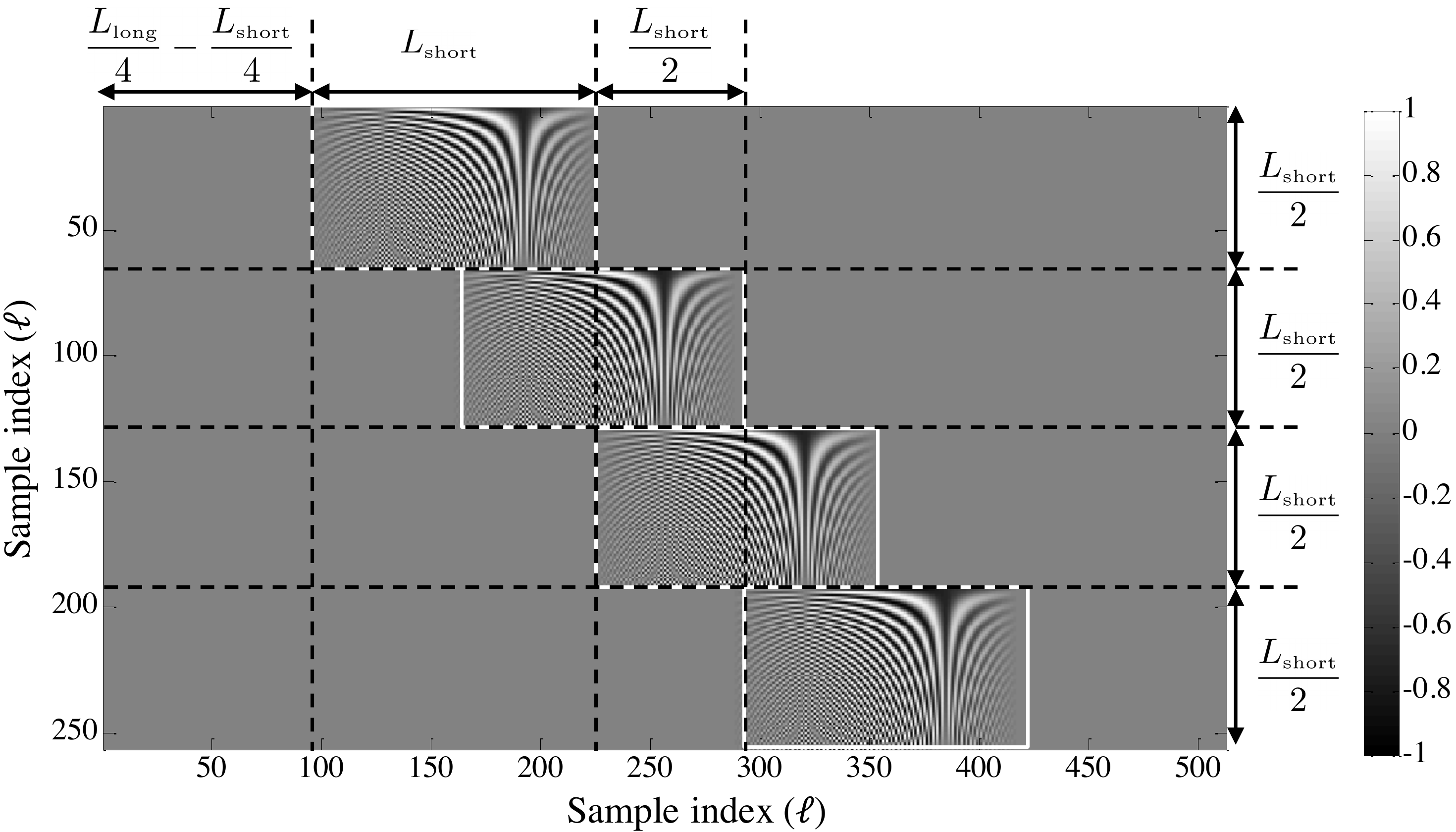}  
  \caption{
  Analysis matrix of short window when $L_{\mbox{\tiny long}}=512$ and $L_{\mbox{\tiny short}}=128$. Four white boxes denote $\bm{ \mathrm{C} }_{\mbox{\tiny short}} \mbox{diag} \left[ \bm{ \mathrm{w} }^{s} \right]$.
  }
  \label{fig:shortmatrix}
\end{figure}

As an objective function for the training of DNN parameters, the following mean-absolute-error (MAE) is often used: 
$\mathcal{J}_{\theta} ^{\mbox{\scriptsize WA}} = \frac{1}{T}\sum_{t=1}^T\lVert\bm{\mathrm{s}}_t - \hat{\bm{\mathrm{s}}}_t \rVert _1,$ 
where $\lVert \cdot \rVert_1$ means the $L_1$ norm, and $\theta = \{ \theta_A, \theta_{M,1}, \theta_{M,2}, \theta_{M,3}, \theta_{M,4} \}$. Since $\mathcal{J}_{\theta} ^{\mbox{\scriptsize WA}}$ independently evaluates the estimated accuracy of $\bm{\mathrm{s}}_t$ for each $t$, it is better for the length of $\bm{\mathrm{s}}_t $ in all $t$ be the same for computational efficiency even though $L$ is not the same in each $t$. 
To satisfy this constraint, the size of the analysis matrices of ``long'' $ \bm{ \mathrm{M} }_{1}$ and ``short'' $ \bm{ \mathrm{M} }_{3}$ must be the same. 
To achieve this, we design $ \bm{ \mathrm{M} }_{3}$ to use the ``short'' $L_{\mbox{\tiny long}} / L_{\mbox{\tiny short}}$ times consecutively. 
Namely, we connect $L_{\mbox{\tiny long}} / L_{\mbox{\tiny short}}$ analysis matrices of the ``short'' window in the row direction, as shown in Fig. \ref{fig:shortmatrix}. 
Then, the analysis matrix outputs the connected $L_{\mbox{\tiny long}} / L_{\mbox{\tiny short}}$ MDCT spectra, 
which are analyzed by the ``short'' window with half-overwrap.

The details of the implementation of each analysis matrix are as follows:
\begin{align}
\bm{ \mathrm{M} }_{1} &= \bm{ \mathrm{C} }_{\mbox{\tiny long}} \mbox{diag} \left[  \bm{ \mathrm{w} }^{l}  \right],\\
\bm{ \mathrm{M} }_{2} &= \bm{ \mathrm{C} }_{\mbox{\tiny long}} \mbox{diag} \left[ \left( \bm{ \mathrm{w} }_{1 }^{l}, \bm{1}, \bm{ \mathrm{w} }_{2}^{s}, \bm{0} \right) \right],\\
\bm{ \mathrm{M} }_{3} \left( \mathcal{I}_{C,h} , \mathcal{I}_{R,h}  \right) 
									&= \bm{ \mathrm{C} }_{\mbox{\tiny short}} \mbox{diag} \left[ \bm{ \mathrm{w} }^{s} \right] \label{eq:short_MDCT_mat}, \\
\bm{ \mathrm{M} }_{4} &= \bm{ \mathrm{C} }_{\mbox{\tiny long}} \mbox{diag} \left[ \left( \bm{0}, \bm{ \mathrm{w} }_{1}^{s}, \bm{1}, \bm{\mathrm{w} }_{2}^{l} \right) \right],
\end{align}
where $\bm{ \mathrm{C} }_{\mbox{\tiny long}}$ and $\bm{ \mathrm{C} }_{\mbox{\tiny short}}$ is the MDCT matrix with $L_{\mbox{\tiny long}} $ and $ L_{\mbox{\tiny short}}$, respectively. 
The ``long'' and ``short'' windows are $\bm{ \mathrm{w} }^{l}$ and $\bm{ \mathrm{w} }^{s}$, respectively, and $\bm{ \mathrm{w} }_{1 }^{l}$ and $\bm{ \mathrm{w} }_{2}^{l}$ denote the first and later half of $\bm{ \mathrm{w} }^{l}$, respectively. 
The vectors $\bm{1}$ and $\bm{0}$ are one/zero vectors with $L_{\mbox{\tiny long}} / 4 - L_{\mbox{\tiny short}} / 4$, respectively, and $\mathcal{I}_{C,h}$ and $ \mathcal{I}_{R,h}$ denote the indexes of a matrix with $ h \in \{ 0,..., L_{\mbox{\tiny long}} / L_{\mbox{\tiny short}}-1 \}$ as follows:
\begin{align}
\mathcal{I}_{C,h} &= \left[ 1 : \frac{L_{\mbox{\tiny short}}}{2} \right] + h\frac{L_{\mbox{\tiny short}}}{2},
\label{eq:short_c_shift}
\\
\mathcal{I}_{R,h} &= \left[ 1 : L_{\mbox{\tiny short}} \right] + \frac{ L_{\mbox{\tiny long}} }{ 4 } - \frac{ L_{\mbox{\tiny short}} }{ 4 } + h\frac{ L_{\mbox{\tiny short}} }{ 2 }.
\label{eq:short_r_shift}
\end{align}

\section{Experiments}
\label{sec:exp}

\subsection{Experimental setup}

\subsubsection{Proposed and comparison methods}

We tested $L_{\mbox{\tiny long}} =512$ and $ L_{\mbox{\tiny short}} = 128$. 
Bi-directional long short-time memory (BLSTM) with two 512-unit layers was used as $\mathcal{M}^{j}$ and $\mathcal{P}$. 
Since the MDCT-spectrum is not shift invariant, we used the modified complex lapped transform (MCLT) spectrum \cite{MCLT} as the input feature of $\mathcal{M}^{j}$;
$\bm{\phi}_{j,t}$ was calculate as the before/after $R=5$ frame concatenated the log-amplitude-MCLT spectrum with $j$-th window length.
The $\bm{\phi}_{1,t}$ was used as the input feature of $\mathcal{A}$ and $\tau = 10^{-4}$ was used as the temperature parameter. 
The rectified linear unit and sigmoid function were used as the activation functions of the first and output layer, respectively. We also used the following two pre-trainings and one fine-tuning; 
(i) $\mathcal{M}^{1}$ and $\mathcal{M}^{3}$ were trained using only ``long'' and ``short'' windows, and $\mathcal{M}^{2}$ and $\mathcal{M}^{4}$ were trained alternately using ``start'' and ``stop'' windows. 
The objective function was $\mathcal{J}_{\theta} ^{\mbox{\scriptsize WA}}$. 
(ii) The $\mathcal{P}$ was trained to minimize
\begin{equation}
\mathcal{J}_{\theta} ^{\mbox{\scriptsize AWS}}  = 
\frac{1}{T} \sum_{t=1}^{T} \sum_{i=1} ^{2} p( a_{i,t} = 1 ) \ln \frac{ p( a_{i,t} = 1 ) }{ q( a_{i,t} = 1 ) }.
\end{equation}
Here, 
$p( a_{1,t} = 1 ) = \bm{e}_{\mbox{\tiny long}, t} / (\bm{e}_{\mbox{\tiny long}, t} + \bm{e}_{\mbox{\tiny short}, t})$ and 
$p( a_{2,t} = 1 ) = \bm{e}_{\mbox{\tiny short}, t} / (\bm{e}_{\mbox{\tiny long}, t} + \bm{e}_{\mbox{\tiny short}, t})$, where
$\bm{e}_{\mbox{\tiny long}, t}$
and
$\bm{e}_{\mbox{\tiny short}, t}$
are 
$\lVert \bm{\mathrm{s}}_{t+1} - \hat{\bm{\mathrm{s}}}_{t+1} \rVert _1$
when using the long and short window, respectively. 
The $q( a_{1,t} = 1 )$ and $q( a_{2,t} = 1 )$ were the outputs of the softmax function of $\mathcal{A}$ instead of the Gumbel-softmax. 
(iii) The $\mathcal{M}^{j}$ and $\mathcal{A}$ were fine-tuned to minimize the following objective function simultaneously:
\begin{equation}
\theta \gets \argmin_{\theta} \left( \mathcal{J}_{\theta} ^{\mbox{\scriptsize WA}} + \lambda \mathcal{J}_{\theta} ^{\mbox{\scriptsize AWS}}  \right),
\end{equation}
where $\lambda = 0.1$.

To investigate the effectiveness of AWS, the proposed method was compared with fix-resolution T-F transforms, {\it i.e.}, the STFT with size 512 points and the MDCT with  $L=512$ and $L=128$. 
The same BLSTM architecture as $\mathcal{M}^{j}$ was used for each method. 
Before/after 5 frames concatenated log amplitude STFT spectrum was used as the input feature for STFT, and 
MCLT-based acoustic features $\bm{\phi}_{1,t} $ and $\bm{\phi}_{3,t} $ were used as that of the MDCTs, respectively. 
Each method was trained for minimizing $\mathcal{J}_{\theta} ^{\mbox{\scriptsize WA}} $.

\subsubsection{Datasets and training setup}

The Wall Street Journal (WSJ-0) corpus and noise dataset CHiME-3 were used as the training dataset. 
The WSJ-0 dataset consisted of 14633 utterances. 
CHiME-3 consisted of four types of background noise: \textit{cafes}, \textit{street junctions}, \textit{public transport (buses)}, and \textit{pedestrian areas} \cite{CHiME}. 
The noisy signals was formed by mixing clean speech utterances with the noise at signal-to-noise ratio (SNR) levels of -6 to 12 dB. 
As the test datasets, 400 utterances randomly selected from the TIMIT corpus were used for the target-source dataset, 
four types of ambient noise {\it F16}, {\it factory 1}, {\it M109}, and {\it Machinegun} from the NOISEX92 dataset were used as the noise dataset. 

The training schedule was designed based on \cite{Erdogan_2018_INTERSPEECH}.
We defined an epoch as having 1k utterances and train with a minibatch of 5 utterances. 
We fixed the learning rate for the initial 100 epochs and decreased it linearly between 100--300 epochs down to a factor of 100 using Adam which was started with a learning rate of $10^{-3}$. 
We also used annealed dropout \cite{aneel_dr} for BLSTM layers, where we started with an initial dropout rate of 0.5 and reduce it linearly after 50 epochs. 
We always concluded training after 300 epochs.

\subsection{Objective experiment}

\begin{table}[ttt]
  \centering
\caption{SDR improvement.}
  \vspace{5pt}
Input SNR: -6 dB\\
  \vspace{3pt}
\small
  \begin{tabular}{l|cccc} \Hline
Method			&	F16		&	Fact. 1	& M109		& Machinegun	\\ \Hline 
STFT			& 7.91   		& 8.13   		& 10.42  		& 12.50		\\ 
MDCT ($L=512$)	& 6.74   		& 7.80   		& 11.69   		& 14.23		\\
MDCT ($L=128$)	& 8.24  		& 9.26   		& 11.60   		& 13.54		\\
Proposed			& {\bf 8.39}	& {\bf 9.29}  	& {\bf 11.78}  	& {\bf 14.28} \\	 \Hline   
  \end{tabular}\\
  \vspace{5pt}
Input SNR: 6 dB\\
  \vspace{3pt}
\small
  \begin{tabular}{l|cccc} \Hline
Method			& F16			& Fact. 1			& M109		& Machinegun	\\ \Hline 
STFT			& {\bf 5.97}  		& 6.10			& 9.12   		& 4.13		\\ 
MDCT ($L=512$)	& 4.79   			& 5.79   			& 8.92   		& 7.98		\\
MDCT ($L=128$)	& 5.92   			& 5.86   			& 7.76   		& {\bf 9.54} 	\\
Proposed			& 5.77   			& {\bf 6.16}	  	& {\bf 9.17}  	& 8.63 \\	 \Hline   
  \end{tabular}\\
  \label{tbl:result}
\end{table}

The speech enhancement performance of the proposed method was compared with those of the conventional methods using SDR-improvement. 
Two input SNR conditions, -6 and 6 dB, were tested. 
Table \ref{tbl:result} shows the evaluation results. 
Under most of input SNR and noise conditions, the proposed method outperformed conventional methods, {\it i.e.} fixed-frequency transforms. 
Although some scores of MDCT ($L=512$) were lower than that of the STFT and MDCT ($L=128$), the proposed method outperformed both methods. 
These results indicate that to locally use a short instead of long window is effective. 
SDRs of STFT were higher than that of MDCT ($L=512$) in some conditions, thus the STFT maybe more effective depending on noise type.
Fortunately, AWS can be used for not only the MDCT but also other T-F transforms including the STFT.
Thus, incorporating the trainable AWS into the STFT will improve speech enhancement performance in the STFT-domain.

\begin{figure}[ttt]
  \centering
\includegraphics[width=80mm,clip]{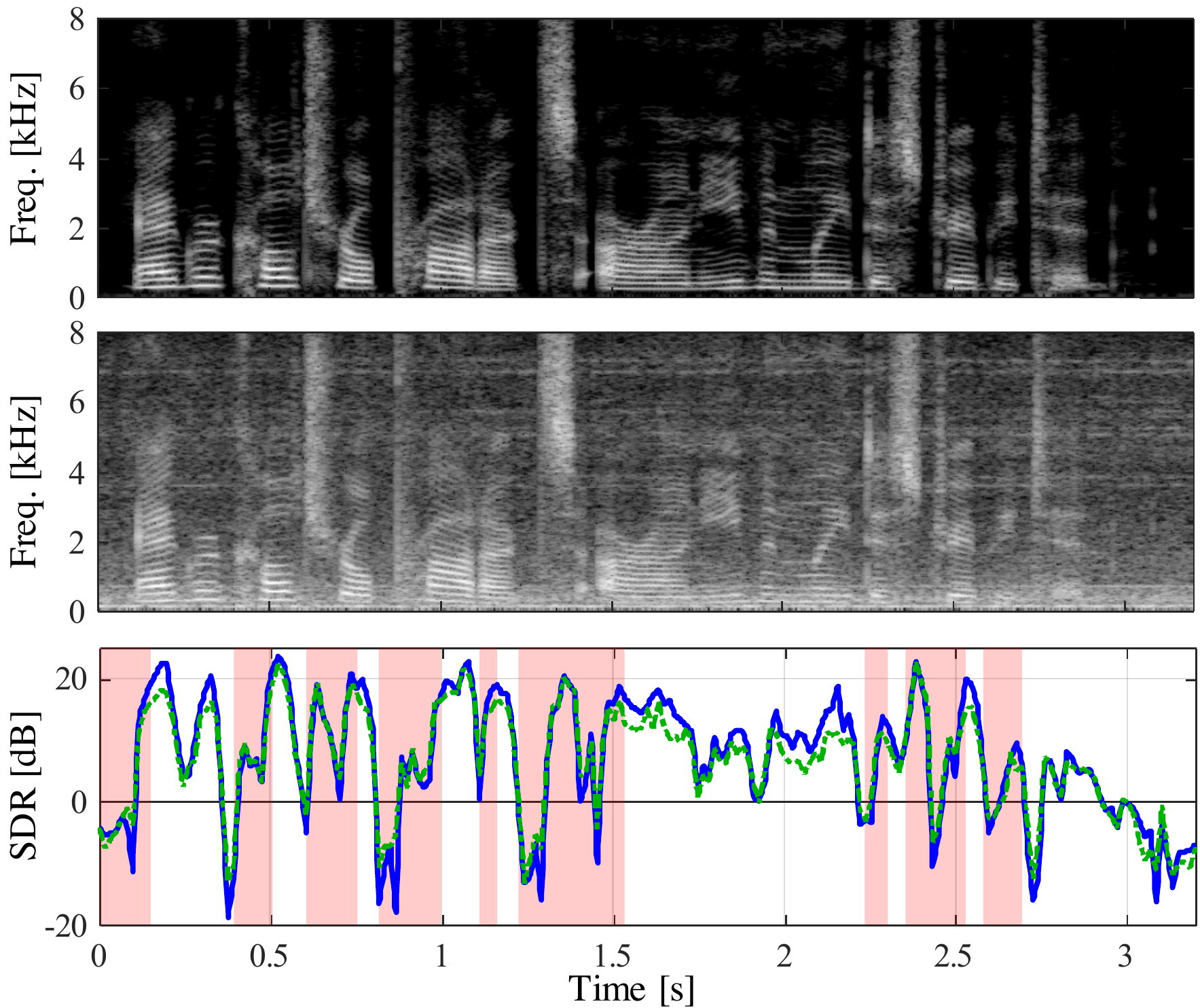}  
  \caption{
Top and middle figures show spectrograms of clean and noisy speech, respectively. 
Bottom figure shows segmental SDRs of estimated signal with 
long (blue) and short (green dotted) window, respectively. Pink area denotes $t$s at which short window was selected.
}
  \label{fig:result_example}
\end{figure}

Figure \ref{fig:result_example} shows an example of the AWS of the proposed method. 
When a ``long'' window was used in all $t$s, the segmental SDRs were higher at around 0.3, and 2.0 sec, namely, the spectrum remained stationary or varied slowly. 
On the other hand, when a ``short'' window was used in all $t$s, the segmental SDRs were higher at around 0.8 and 1.4 sec, namely, significant change point of phoneme.
The proposed method selected the better window when a clear difference appeared in the segmental SDR. 
This may be a reason the proposed method outperformed the fixed-resolution T-F transforms.

\section{Conclusions}
\label{sec:cncl}

We proposed a trainable AWS method and applied it to the MDCT-domain speech enhancement. 
AWS is incorporated into the speech enhancement procedure and the parameters for manipulating each window are estimated using a DNN. 
The experimental results indicate that the proposed method outperformed the fixed-resolution T-F transforms.
Thus, we conclude that the proposed method can be effective for speech enhancement.

In the experiments, the proposed method was not compared with trainable T-F transforms because trainable AWS with these methods is not an antithetical concept. 
Thus, we plan to develop a more flexible trainable T-F transform; simultaneous optimization of trainable T-F transforms and AWS.

\clearpage
\bibliographystyle{IEEEbib}
\bibliography{refs}

\begin{thebibliography}{99}
\vspace{-3mm}
%
\bibitem{Wang_2018} 
D.~L.~Wang and J.~Chen,
``Supervised Speech Separation Based on Deep Learning: An Overview,'' 
{\it IEEE/ACM Trans. on Audio, Speech, and Lang. Process.,} 2018.
\bibitem{Erdogan_2015} 
H.~Erdogan, J.~R.~Hershey, S.~Watanabe, and J.~L.~Roux,
``Phase-Sensitive and Recognition-Boosted Speech Separation using Deep Recurrent Neural Networks,'' 
{\it Proc. of Int. Conf. on Acoust., Speech, and Signal Process. (ICASSP)}, 2015.
\bibitem{Hershey_2016}
J.~R.~Hershey, Z.~Chen, J.~L.~Roux, and S.~Watanabe,
``Deep Clustering: Discriminative Embeddings for Segmentation and Separation,''
{\it Proc. of Int. Conf. on Acoust., Speech, and Signal Process. (ICASSP)}, 2016.
\bibitem{Kolbak_2017} M.~Kolbak, D.~Yu, Z.~H.~Tan, and J.~Jensen, 
``Multi-talker Speech Separation with Utterance-level Permutation Invariant Training of Deep Recurrent Neural Networks,''
{\it IEEE/ACM Trans. on Audio, Speech, and Lang. Process.,} 2017.
\bibitem{Koizumi_ICASSP_2017} Y.~Koizumi, K.~Niwa, Y.~Hioka, K.~Kobayashi and Y.~Haneda,
``DNN-based Source Enhancement Self-Optimized by Reinforcement Learning using Sound Quality Measurements,''
{\it Proc. of Int. Conf. on Acoust., Speech, and Signal Process. (ICASSP)}, 2017.
\bibitem{Koizumi_TASL_2018} Y.~Koizumi, K.~Niwa, Y.~Hioka, K.~Kobayashi and Y.~Haneda,
``DNN-based Source Enhancement to Increase Objective Sound Quality Assessment,''
{\it IEEE/ACM Trans. on Audio, Speech, and Lang. Process.,}, 2018.


\bibitem{Will_cIRM_2016} D.~S.~Williamson, Y.~Wang and D.~L.~Wang,
``Complex Ratio Masking for Monaural Speech Separation,''
{\it IEEE/ACM Trans. on Audio, Speech, and Lang. Process.,}, pp.483--492, 2016.
\bibitem{GL_1984} D.~W.~Griffin and J.~S.~Lim, 
``Signal Estimation from Modified Short-Time Fourier Transform,''
{\it IEEE Trans. on Audio, Speech, and Signal Process.}, 1984.
\bibitem{Yatabe_2018} K.~Yatabe, Y.~Masuyama and Y.~Oikawa,
``Rectified Linear Unit Can Assist Griffin-Lim Phase Recovery,''
{\it Proc. of Int. Workshop on Acoustic Signal Enhancement (IWAENC)}, 2018.
\bibitem{Wakabayashi_2018}
Y.~Wakabayashi, T.~Fukumori, M.~Nakayama, T.~Nishiura, and Y.~Yamashita,
``Single-Channel Speech Enhancement with Phase Reconstruction Based on Phase Distortion Averaging,''
{\it IEEE/ACM Trans. on Audio, Speech, and Lang. Process.,} pp.1559--1569, 2018.
\bibitem{Masuyama_2018} Y.~Masuyama, K.~Yatabe and Y.~Oikawa,
``Model-based Phase Eecovery of Spectrograms via Optimization on Riemannian Manifolds,''
{\it Proc. of Int. Workshop on Acoustic Signal Enhancement (IWAENC)}, 2018.

\bibitem{PhaseNet} N.~Takahashi, P.~Agrawal, N.~Goswami, and Y.~Mitsufuji,
``PhaseNet: Discretized Phase Modeling with Deep Neural Networks for Audio Source Separation,''
{\it Proc. Interspeech}, 2018.
\bibitem{GAN_phase} K.~Oyamada, H.~Kameoka, T.~Kaneko, K.~Tanaka, N.~Hojo, and H.~Ando,
``Generative Adversarial Network-based Approach to Signal Reconstruction from Magnitude Spectrograms,''
{\it Proc. of European Signal Processing Conf. (EUSIPCO)}, 2018.
\bibitem{Phasebook} J.~{Le Roux}, G.~Wichen, A.~Watanabe, A.~Sarroff, and J.~R.~Hershey,
``Phasebook and Friends: Leveraging Discrete Representations for Source Separation,''
{\it arXiv preprint,} arXiv:1810.01395, 2018.
\bibitem{DeGLI} 
Y.~Masuyama, K.~Yatabe, Y.~Koizumi, N.~Harada, Y.~Oikawa,
``Deep Griffin--Lim Iteration,''
{\it Proc. of Int. Conf. on Acoust., Speech, and Signal Process. (ICASSP)}, 2019.

\bibitem{MDCT_01} J.~P.~Prince and A.~B.~Bradley,
``Analysis/Synthesis Filter Bank Design Based on Time Domain Aliasing Cancellation,''
{\it IEEE/ACM Trans. on Audio, Speech, and Lang. Process.,} pp.1153--1161, 1986.
\bibitem{Kuech_2007} F.~Keuch and B.~Elder,
``Aliasing Reduction for Modified Discrete Cosine Transform Domain Filtering and Its Application to Speech Enhancement,''
{\it Proc. of IEEE Workshop on Applications of Signal Process. to Audio and Acoust. (WASPAA)}, 2007.
\bibitem{Koizumi_ICASSP_2018} Y.~Koizumi, N.~Harada, Y.~Haneda, Y.~Hioka, and K.~Kobayashi,
``End-to-End Sound Source Enhancement using Deep Neural Network in the Modified Discrete Cosine Transform Domain,''
{\it Proc. of Int. Conf. on Acoust., Speech, and Signal Process. (ICASSP)}, 2018.
\bibitem{Venkataramani_2017} 
S.~Venkataramani, J.~Casebeer, and P.~Smaragdis,
``End-to-end Source Separation with Adaptive Front-Ends,''
{\it Proc. of Asilomar Conf. on Signals, Systems and Computers (ACSSC)}, 2018.
\bibitem{Venkataramani_2018} 
S.~Venkataramani, and P.~Smaragdis,
``End-to-end Networks for Supervised Single-channel Speech Separation,''
{\it arXiv preprint,} arXiv:1705.02514, 2018.
\bibitem{Wichern_2018} 
G.~Wichern, and J.~Le Roux,
``Phase Reconstruction with Learned Time-Frequency Representations for Single-Channel Speech Separation,''
{\it Proc. of Int. Workshop on Acoustic Signal Enhancement (IWAENC)}, 2018.
\bibitem{Luo_2018} 
Y.~Luo, and N.~Mesgarani,
``TasNet: Surpassing Ideal Time-Frequency Masking for Speech Separation,''
{\it arXiv preprint,} arXiv:1809.07454, 2018.
\bibitem{Takeuchi_2019} D.~Takeuchi, K.~Yatabe, Y.~Koizumi, N.~Harada, and Y.~Oikawa,
``Data-Driven Design of Perfect Reconstruction Filterbank for DNN-based Sound Source Enhancement,''
{\it Proc. of Int. Conf. on Acoust., Speech, and Signal Process. (ICASSP)}, 2019.
\bibitem{Mochizuki_1994} T.~Mochizuki, 
``Perfect Reconstruction Conditions for Adaptive Blocksize MDCT,''
{\it IEICE Trans. on Fund. of Elect., Comm. and Computer Sciences}, 1994.
\bibitem{AWS} V.~Britanak, and K.~R.~Rao,
``Cosine-/Sine-Modulated Filter Banks, General Properties, Fast Algorithms and Integer Approximations,''
{\it Springer}, 2018
\bibitem{mp3} ISO/IEC 11172-3:1993
``Coding of Moving Pictures and Associated Audio for Digital Storage Media at up to about 1,5 Mbit/s--Part 3: Audio,''
1993.
\bibitem{gumbel-softmax}
E.~Jang, S.~Gu, and B.~Poole,
``Categorical Reparameterization with Gumbel-Softmax,''
{\it Proc. of Int. Conf. on Learning Representations, (ICLR)}, 2017.
\bibitem{MCLT} H.~Malvar, 
``A Modulated Complex Lapped Transform and its Applications to Audio Processing,''
{\it Proc. of Int. Conf. on Acoust., Speech, and Signal Process. (ICASSP)}, 1999.
\bibitem{CHiME} J. Barker, R. Marxer, E. Vincent and S. Watanabe,
``The third `CHiME' Speech Separation and Recognition Challenge: Dataset, Task and Baseline,''
{\it Proc. of IEEE Automatic Speech Recognition and Understanding Workshop (ASRU),} 2015.



\bibitem{Erdogan_2018_INTERSPEECH} 
H.~Erdogan, and T.~Yoshioka,
``Investigations on Data Augmentation and Loss Functions for Deep Learning Based Speech-Background Separation,''
{\it Proc. of Interspeech}, 2018. 
\bibitem{aneel_dr} 
S.~J.~Rennie, V. Goel, and S. Thomas, 
``Annealed dropout training of deep networks''
{\it Proc. of Spoken Language Technology Workshop (SLT)}, 2014.


\end{thebibliography}

\end{document}